\documentclass[submission,copyright,creativecommons]{eptcs}
\providecommand{\event}{F-IDE 2021} 

\usepackage[english]{babel}
\usepackage{amssymb}
\usepackage{caption}
\usepackage{float}
\usepackage{graphicx}
\usepackage{hyperref}
\usepackage{listings}
\usepackage{xcolor}

\hypersetup{
  colorlinks=true,
  pdflang={English}}

\definecolor{keyword}{rgb}{0,0,0.6}
\definecolor{type}{rgb}{0.6,0,0.6}
\definecolor{identifier}{rgb}{0,0.4,0}

\lstdefinelanguage{dezyne}{
  basicstyle=\ttfamily\footnotesize,
  keywordstyle=[0]\color{keyword}\bfseries,
  keywordstyle=[1]\color{type}\bfseries,
  otherkeywords={<=>},
  keywords=[0]{behaviour,component,in,interface,on,out,provides,requires,system},
  keywords=[1]{bool,enum,subint,void},
}

\title{Dezyne: Paving the Way to\\ Practical Formal Software Engineering}
\author{Rutger van Beusekom, Bert de Jonge, Paul Hoogendijk, Jan Nieuwenhuizen
\institute{Verum Software Tools B.V., The Netherlands}
\email{\{rutger.van.beusekom,bert.de.jonge,paul.hoogendijk,jan.nieuwenhuizen\}@verum.com}
}
\def\titlerunning{Dezyne: Paving the Way to Practical Formal Software Engineering}
\def\authorrunning{R. van Beusekom, B. de Jonge, P. Hoogendijk \& J. Nieuwenhuizen}
\def\copyrightholders{Van Beusekom, De Jonge, Hoogendijk \& Nieuwenhuizen}

\begin{document}
\maketitle

\begin{abstract}
Designing software that controls industrial equipment is challenging,
especially due to its inherent concurrent nature. Testing this kind of
event driven control software is difficult and, due to the large
number of possible execution scenarios only a low dynamic test
coverage is achieved in practice. This in turn is undesirable due to
the high cost of software failure for this type of equipment.

In this paper we describe the Dezyne language and tooling; Dezyne is a
programming language aimed at software engineers designing large
industrial control software. We discuss its underlying two layered and
compositional approach that enables reaping the benefits of Formal
Methods, hereby strongly supporting guiding principles of software
engineering. The core of Dezyne uses the mCRL2 language and
model-checker (Jan Friso Groote et al.) to verify the correctness and
completeness of all possible execution scenarios.

The IDE of Dezyne is based on the Language Server Protocol allowing a
smooth integration with e.g., Visual Studio Code, and Emacs, extended
with several automatically generated interactive graphical views. We
report on the introduction of Dezyne and its predecessor at several
large high-tech equipment manufacturers resulting in a decrease of
software developing time and a major decrease of reported field
defects.
\end{abstract}

\section{Introduction}

Designing software that controls industrial equipment is challenging.
The complexity of such equipment is ever growing and so is the demand
on its software. The software must control and monitor many processes
in parallel. Designing the regular control flow of the processing
steps is already complex, but many exceptional conditions may occur
which disrupt the regular control flow and must be handled
appropriately.

Testing event driven control software is difficult.  Testing time on
hardware is expensive so often software simulators on various levels
of (hardware) details are created to mitigate these costs. However,
the ultimate test is still the execution on the real hardware; errors
found during the final testing are very costly and time consuming to
fix.  Writing test cases for (parts of) the software is difficult
since, as explained earlier, many different execution scenarios have
to be considered which depend on different timing conditions. As a
result, only a small percentage of the possible execution scenarios is
effectively tested in practice.

This kind of high-tech equipment is extremely expensive, for instance
the Extreme Ultraviolet (EUV) lithography machine of ASML costs around
\$120 million dollar.  Manufacturers of such kind of equipment are
required to deliver a high uptime of their machines. In some cases,
the manufacturers are penalized when a machine does not reach the
agreed availability. As a result, down time of these machines as the
result of software failure must be avoided at all cost.

\section{Formal Methods}
As stated in the previous section, designing and testing control
software of industrial high-tech equipment in a traditional fashion
has serious short comings, while on the other hand the
cost-of-non-quality is extremely high.

Formal Methods have been around for many decades and have been shown
to deliver reliable and safe software.  However, Formal Methods come
with their own challenges when applied to industrially sized systems
and processes.  The tools and their associated languages require
Formal Methods specialists.  Furthermore, a translation must be made from
the (informal) requirements to some description written in a Formal
Methods Language.  Typically, it is challenging to validate that the
intended requirements are correctly captured or are that the
requirements are appropriate in the first place.

The artifacts of the formalization process are typically used by
Formal Methods toolsets in proving the consistency of the
requirements and any of the derived refinements of the specifications.
However, ultimately we need to obtain running code on the hardware for
which the specified properties provably hold.

The total state space of any industrial size machine controlling
application is vast and cannot be handled by any Formal Methods tool
in its entirety.  A practical way to handle this is to take a
compositional approach: at a certain level one should be able to
abstract away from details which are not relevant for the next
level. This way the state-explosion problem can be avoided.

\section{Dezyne}
Verum has developed Dezyne \cite{BeusekomGHHWWW17}. Dezyne is a
programming language aimed at software engineers designing large
industrial control software.  The language is designed to have a very
low barrier to entry for regular software engineers. The Dezyne
language consists of two declarative language constructs: the
so-called ``guarded'' statement and the ``on'' statement for
specifying the states and trigger events of a state machine,
respectively. Furthermore, the Dezyne language has de facto standard
imperative language constructs as present in regular languages like C
or Java: variable declaration, assignment, function declaration,
function call, and control flow constructs like if-then-else.

\subsection{The Component Model of Dezyne}
Dezyne has the notion of three models: interfaces, components, and
systems.

An interface model contains the definition of a set of events. In
Dezyne events are implemented as function calls.  Each event
declaration has a direction, ``in'' or ``out'', and a type signature. The
type signature of an event specifies the data parameters carried by an
event, and the type of the return value of an event.  The allowed
return types are: ``void'', ``boolean'', enumeration types, and
integer-range types.  The types of the data parameters can be any
arbitrary type.  This is an example of event declarations of an
interface:
\begin{lstlisting}[language=dezyne]
interface IDevice {
  in void turnon();
  in void turnoff();
}
\end{lstlisting}
An interface also has a behaviour section. The behaviour of an
interface prescribes the order in which events are expected to
occur. A behaviour can specify additional state variables for keeping
track of the state of the interface. Using the simulator of the Dezyne
tooling, the user can produce sequence diagrams depicting allowed
sequences of events for that interface.

An interface plays the role of a contract between
components: it specifies how a client component via its requires port
should interact with the underlying component providing the required
functionality via its provides port.  This is an example of a simple
interface that consists of two events ``turnon'' and ``turnoff''. The
behaviour of the interface specifies that the events should be used
alternatingly:
\begin{lstlisting}[language=dezyne]
interface IDevice {
  in void turnon();
  in void turnoff();
  behaviour {
    enum State {On, Off};
    State s = State.Off;
    [s.Off] {
      on turnon: s = State.On;
      on turnoff: illegal;
    }
    [s.On] {
      on turnon: illegal;
      on turnoff: s = State.Off;
    }
  }
}
\end{lstlisting}
Note that the statement ``\verb+on turnoff: illegal;+'' defines that
the event ``turnoff'' is not allowed in that state.

A component model and a system model start by declaring their
provides
ports and requires ports.  A port is a instance of an interface.

A component model also specifies a behaviour, i.e.  the actual
implementation of the component. The behaviour describes how a
component implements each of the events of its provides ports hereby
possibly using some of its requires ports. A behaviour can specify
additional variables to keep track of the state of the interaction
between the component and its ports. In this way the user specifies a
finite state-machine describing the implementation of the component.
This is an example of a component model:
\begin{lstlisting}[language=dezyne]
component Fork {
  provides IDevice p;
  requires IDevice r0;
  requires IDevice r1;
  behaviour {
    enum State {On, Off};
    State s = State.Off;
    [s.Off] {
      on p.turnon(): {r0.turnon(); r1.turnon(); s = State.On;}
    }
    [s.On] {
      on p.turnoff(): {r0.turnoff(); r1.turnoff(); s = State.Off;}
    }
  }
}
\end{lstlisting}

\begin{figure}[H]
  \centering
  \includegraphics[scale=0.6]{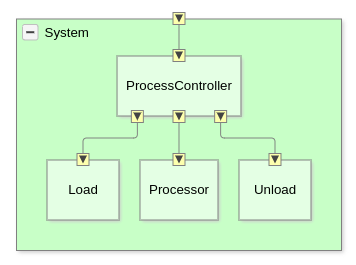}
  \caption{Dezyne system diagram}
  \label{figure:system}
\end{figure}

Next to the provides and requires ports, a system model lists
component instances and the bindings between the ports of the
component instances and the external ports of the system. These
bindings determine the routing, i.e. call graph, of the events of the
external ports of the systems and the component instances.
This is an example of a system model:
\begin{lstlisting}[language=dezyne]
component System {
  provides IControl cntrl;

  system {
    ProcessController controller;
    Load loader;
    Processor processor;
    Unload unloader;

    cntrl <=> controller.cntrl;
    controller.load <=> loader.load;
    controller.process <=> processor.process;
    controller.unload <=> unloader.unload;
  }
}
\end{lstlisting}
The Dezyne tools can generate a diagram from the system model as show
in figure \ref{figure:system}.

\noindent
Section \ref{section:ide} explains in more detail how these diagrams
are generated.  Note that each connection in the diagram corresponds
to a binding statement \verb+<=>+ in the system model text.

\subsection{Verifying Dezyne Models}
The Dezyne language is statically checked on the use of syntax, types
and an extensive list of well-formedness properties\footnote{Among
others: binding pairs of provides and requires; non circular binding
of ports; allowed component triggers are ``in'' events of provides ports
and ``out'' events of requires ports}.

During the verification phase the Dezyne verification engine checks
interface and component models.  The verification engine translates
the Dezyne model to an mCRL2 description
\cite{mcrl2book,mcrl2tool,BeusekomGHHWWW17}.  Next, the mCRL2 tool set
is used to convert the mCRL2 description to an LTS (labeled transition
system). An LTS is a graph where nodes relate to states of the Dezyne
model and each edge is labeled with an event instance. The graph
describes the total state space of the model and all possible
execution scenarios that cover this state space.

Both an interface and a component model are verified for the
absence of deadlocks and livelocks, and whether all integer assignments are
in range by inspecting the generated LTS. Additionally a component
model is verified to correctly interact with its requires ports
as specified by their interfaces. For this, the LTS is checked for the absence
of the ``illegal'' label, where the presence would indicate that either
the component performed an action on one of its requires port which
was disallowed by the corresponding interface behaviour, or that a requires port
performed an event on the component which was disallowed by the component behaviour.

Finally, the compliance of a
component with all of its provides ports is verified by means of the
\emph{ltscompare} tool of mCRL2 using the Failures Refinement preorder relation known from CSP
\cite{hoare1985communicating,roscoebook,mcrl2fdr}.  It is used to verify that
the LTS of the component after hiding all internal, i.e. those events
not observable by the clients using the provides ports, is a
refinement of the LTS of the provides ports, i.e. for component C with
provides port of type I and requires port type J it verifies that:
\begin{equation}
  (C\|J) \! \upharpoonright \! \alpha (I) \, \sqsupseteq_{F} \, I
\end{equation}
where $\sqsupseteq_{F}$ denotes the Failures Refinement relation of
CSP, $\alpha(I)$ denotes the alphabet of process $I$, $\|$ denotes the
parallel composition synchronising on the common alphabet, and
$\upharpoonright$ denotes the projection operator, i.e. the complement
of the hiding operator, i.e. $P\!\upharpoonright\!A =
P\setminus\overline{A}$.

If an error is found a counter example is produced. This is a trace of
events leading up to the state where the inconsistency emerges.

\begin{figure}[ht]
  \centering
  \includegraphics[scale=0.5]{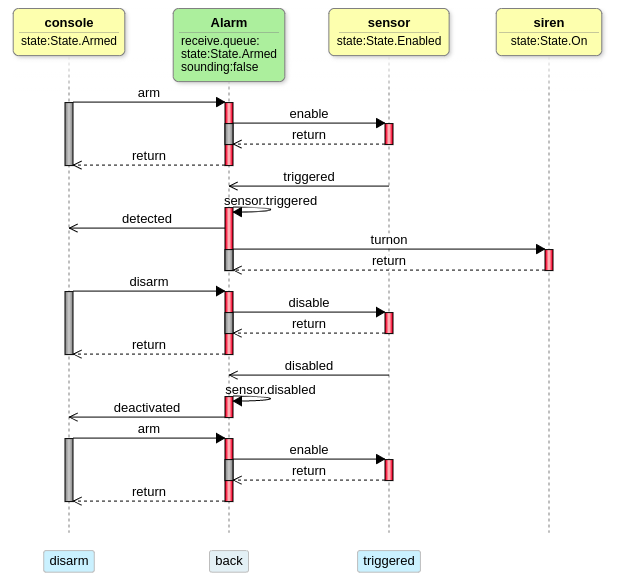}
  \caption{Generated Sequence Diagram with buttons for selecting next
    possible event.}
  \label{figure:seq}
\end{figure}

Next in the Dezyne verification pipeline, the trace is fed to the
Dezyne simulator together with the original Dezyne models. The Dezyne
simulator reconstructs the relevant state and location information
such that the user can relate the counter example back to the Dezyne
model text. The Dezyne simulator outputs an encoding of a
sequence diagram.  This encoding also allows navigating the source code
location of each event occurrence in the sequence diagram.  This
sequence diagram is fed to a graphical engine implemented in
JavaScript using a graphics library for rendering the sequence
diagram. The Dezyne simulator also outputs all possible valid next
events. In the sequence diagram these events are shown as buttons on
the corresponding life lines.  Figure \ref{figure:seq} depicts an
example of a sequence diagram.

\subsection{Two Level Approach}
As described in the previous section, when verifying a Dezyne
model a user does not need to have a knowledge about Formal Methods or the
mCRL2 language: under the hood the Dezyne models are translated to
mCRL2 models which are verified using the mCRL2 tooling and a LTS
analysis tool. If a problem is found, the counter example trace is fed
to the Dezyne simulator that generates an interactive sequence diagram
for the user. The user can use this diagram to interactively find the
root cause. The sequence diagram shows what the component is doing
from the start of the system up to the reported violation. If the user
has diagnosed the problem and has fixed the Dezyne models, they can
rerun the Dezyne verification engine and see whether the issue has
disappeared, and if not, iterate the steps mentioned above until the issue has
been fixed. After which, possibly another issue is found by the
verification engine and the attention is shifted to resolve this issue
until no more issues remain.

\subsection{Generating Verified and Reliable Code}

When the Dezyne models are verified and are correct, the models are
used for generating production quality machine code.  Currently the
Dezyne tooling has code generators for the languages C++, C\#,
JavaScript and GNU Guile Scheme.

The operational semantics of the Dezyne models as specified by the
modeling in mCRL2 \cite{BeusekomGHHWWW17} and the implementation of
Dezyne simulator is designed to capture the intended execution
behaviour of the corresponding generated machine code.  For Dezyne it
is essential that the three embodiments of the semantics of Dezyne are
equivalent. Given that equivalence, it follows that if no problems are
found during verification, we know that all the verified properties
also hold for the generated machine code.

The equivalence between the semantics used during verification and the
simulation is asserted by means of \emph{ltscompare}-ing the trace
equivalence of the underlying LTSses for a large set of test
models. Thus, for each model M of this test set we check:

\begin{equation}
  LTS_V(M) \, \equiv_T \, LTS_S(M)
\end{equation}
where $LTS_V$ denotes the LTS produced by the mCRL2 tooling as used
during verification and $LTS_S$ denotes the LTS produced using the
simulator by calculating the transitive closure of the single step
function of the simulator, and $\equiv_T$ denotes the Trace
Equivalence relation of CSP, i.e. $L_0 \equiv_T L_1$ iff $traces(L_0)
= traces(L_1)$.

The equivalence of $LTS_V(M)$ with the generated code for component M
is verified in the following way. For $LTS_V(M)$ a trace graph cover
is calculated, i.e. a set of traces such that all edges of $LTS_V(M)$
occur in the trace set.  Next to the generated code for component M, a
simulation stub is also generated.  The generated stub processes the
textual representation of a trace: for an inward going event the
corresponding implementation of the event is called. All outward
events are logged and matched with the events of the trace being
processed.  For all traces of the LTS graph cover, the generated code
plus stub is executed.  If all traces of $LTS_V(M)$ are correctly
handled by the generated machine code for component M, we conclude
that
\begin{equation}
  traces(LTS_V(M)) \, \subseteq \, traces(CODE_L(M))
\end{equation}
where $CODE_L(M)$ denotes the generated code and stub in language L for model M,
and $traces$ on code denotes the set of traces the execution of
the code would accept as described above. 

Since $LTS_V(M)$ is complete, i.e. for all stable states, all events
are present, it follows that $CODE_L(M)$ is also complete, and since
the generated code is deterministic, i.e. for a given input event,
only one response is generated, we know that the generated code cannot
exhibit more behaviour as tested by the generated traces. In other
words, it is that case that:
\begin{equation}
  traces(LTS_V(M)) \, \subseteq \, traces(CODE_L(M)) \,\,\,\,\, \Rightarrow \,\,\,\,\, traces(LTS_V(M)) \, = \, traces(CODE_L(M))
\end{equation}
Hence, we verify using \emph{ltscompare} and the running of all generated traces that
\begin{equation}
  LTS_V(M) \, \equiv_T \, LTS_S(M) \, \equiv_T \, CODE_L(M)
  \label{equation:trace-equivalence}
\end{equation}
for all of the model M of the test set, for a given language L.

The above mentioned test set covers all aspects of the Dezyne language
and expected feature interactions. This test capability is also
available for our users to allow them to verify the equivalence
(\ref{equation:trace-equivalence}) of their own (proprietary) set of
models. As a result, the user can assure that if no problems are found
during verification, all the verified properties hold for the
generated machine code for their specific models in their environment,
i.e. target compiler and platform.

\section{Component Based Providing a Compositional Approach}

As described above, a compositional approach to avoid the state
explosion problem is highly desirable to be able to verify large
complex system designs. Dezyne is a component based method: a
component specifies its provides and requires ports where ports are
instances of interfaces. A Dezyne interface has a contract, i.e. its
behaviour, specifying the contract between two components. The
interfaces of the provides ports of a component are the abstraction of
said component. When designing a Dezyne component, a user does not
need to know which implementation is going to provide an interface; it
is sufficient to know the interface.

This provides a natural way of decomposing a larger system into
smaller manageable parts.  When presented with the challenge to devise
certain functionality, one may approach this task by distinguishing
the separate responsibilities of that functionality which be
separately assigned to one or more components.  The responsibility of
a single component starts with its provides ports and ends with its
requires ports. When designing a component one can, through
verification, justly assume that the underlying components implements
the interfaces of its provides ports faithfully.

The approach of interfaces as an abstraction of components not only
allows the engineer to divide and conquer a problem by considering
each responsibility of a component one at the time, but also
is a solution to verify complete systems
compositionally. This is due to an important property of the Failures Refinement
relation of CSP, which allows verifying each individual component to
be correct and therefore knowing that the overall system is also
correct.  It is the case that when two components refine their
provides port interfaces then the compound of two components will also
refine the top level provides port since the Failures Refinement
relation is transitive and the hiding operator is congruent.  We have,
for component C with provides port of type I and requires port of type J, and
component D with provides port of type J and requires port of type K, that:
\begin{equation}
  (C\|J) \! \upharpoonright \! \alpha (I) \, \sqsupseteq_{F} \, I
  \,\,\, \wedge \,\,\, (D\|K) \! \upharpoonright \! \alpha (J) \,
  \sqsupseteq_{F} \, J \,\,\,\,\, \Rightarrow \,\,\,\,\, (C\|D\|K)
  \!\upharpoonright\!\alpha (I) \, \sqsupseteq_{F} \, I
\end{equation}

Hence, for Dezyne, if we have proven that all components are correct, we have proven that
the complete system is correct.

\section{Guiding Principles of Software Engineering}

The development of the Dezyne methodology and language design are
guided by the fundamental software engineering principles as mentioned
in e.g. \cite{fundamentals}.  These engineering principles are: rigor
and formality, separation of concerns, modularity, abstraction,
anticipation of change, generality and incrementality. We strive to
have a methodology and language in supporting the user as much as
possible in practicing these engineering principles.  Next we discuss
each of the principles in more detail and how they are reflected in
Dezyne in supporting the user.

\subsubsection*{Rigor and formality}
Following this principle will add precision to and increases
confidence in the outcome of the creative process of constructing
software by expressing requirements and ideas succinctly using a well
understood notation which expresses intent and specifically allows
communicating information.

As mentioned before, the Dezyne language has a rigorously and formally
defined operational semantics having three different but equivalent
embodiments: the modeling in mCRL2, the Dezyne simulator, and
the behaviour of the generated machine code.

For the user defining a Dezyne interface describing the interaction
between two components forces the user to clearly specify the expected
interaction for all possible scenarios thus also for exceptional
cases. The verification engine will find for a Dezyne component whether 
there are missing cases or whether there are states
where events are not handled as they are intended.

\subsubsection*{Separation of concerns}
The principle of separation of concerns is a form of divide and conquer
of the work itself as well as dealing with the complexity of problems
in general. By explicitly distinguishing one aspect from another, we
can avoid having to deal with the compounded problem.

As mentioned before Dezyne is component based and guides the user in
finding a proper decomposition of the problem, i.e.  describing
different aspects in behaviours independently in separate interfaces,
component, and system models.

\subsubsection*{Modularity}
Modularity is a structural and behavioural separation of concerns by
encapsulation\footnote{Information hiding}, abstraction and
(de)composition.

Interfaces represent and encapsulate the entire interactional
conversation between peer components. Components encapsulate all of
the coordination across its ports. Systems encapsulate component instances
and their inter-connections.

\subsubsection*{Incrementality}
The principle of incrementality allows for stepwise refining solutions
to challenges, thereby avoiding the problems associated with a big
bang approach.

Verification and code generation can be used throughout the life cycle
of every model. Also models can be extended aspect by aspect and
hereby support an agile software development approach.

\subsubsection*{Abstraction}
Abstraction allows us to distinguish between key issues and side
issues. The key issues must be tackled early in the development
process and will allow to postpone dealing with the details at the
last responsible moment.

Dezyne interfaces are abstractions of component interactions and hide
component implementation details and as mentioned before are
key in having a compositional approach.

\subsubsection*{Generality}
Awareness of generality allow us to move from a point solution to a
general solution, which in turn caters for avoiding (future) work by using
the general solution.

In Dezyne interfaces represent the client's perspective on the
implementation. Taking the different point solutions under consideration, 
this allows deriving a general interface by capturing the
commonality in the interaction of different clients.                              

\subsubsection*{Anticipation of change}
Change is inevitable, therefore we must both keep track of all of the
artifacts of the creative process, as well as maintain the consistency
across the refinement steps. We must also maintain their malleability.

In Dezyne the language and the verification create the freedom to
evolve any aspect while maintaining the consistency during the entire
development lifecycle. If a change breaks the consistency the
verification engine will report this and the user can take appropriate
action.

\section{Dezyne IDE Based on LSP}\label{section:ide}

The core of the Dezyne tooling consists of the verification engine,
simulator and code generators which are implemented using GNU Guile
Scheme. In order to support a multitude of different IDEs we have
implemented a language server implementing the Language Server
Protocol (LSP). LSP allows clients to remain language-agnostic and
share a single language-specific server implementation. Currently the
Dezyne LSP server provides code completion and code navigation.  The
LSP server shares the parsing part with the core Dezyne tooling and as
such only a single language front-end has to be maintained. When the
Dezyne language is extended, it becomes immediately available for
all IDEs by means of the LSP server.

\begin{figure}[hb]
  \centering
  \includegraphics[scale=0.345]{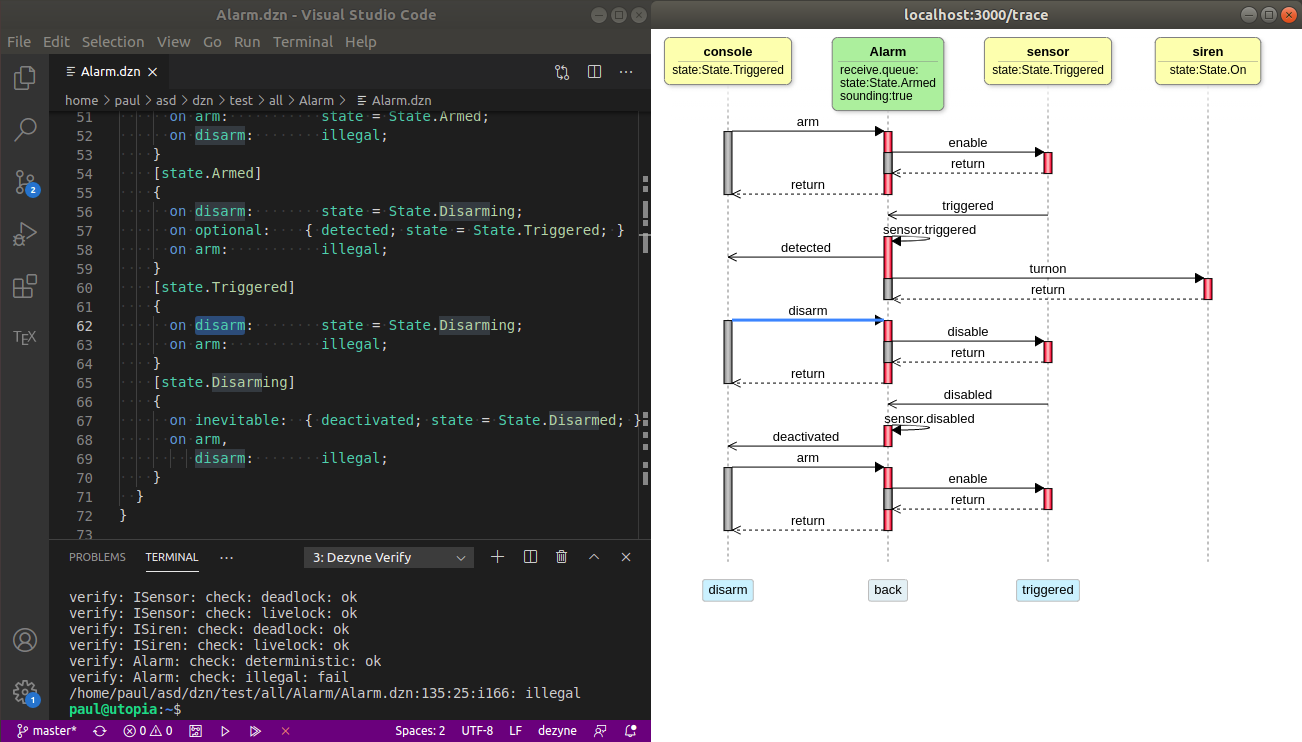}
  \caption{Visual Studio Code with Dezyne LSP and extended with a
    interactive graphical view.}
  \label{figure:vscode}
\end{figure}

Currently there are LSP client implementations of both
Visual Studio Code (\ref{figure:vscode}) and GNU Emacs available. 
Due to the nature of LSP supporting the LSP clients of other
IDEs should involve little work, if any.

LSP is text oriented and currently does not provide any graphical
support. To integrate the interactive Dezyne diagrams, there is a
daemon process running on the machine of the user which among
other things implements a webserver. For all the Dezyne views a
webbrowser is started that connects to this local webserver. The
editor also connects to this webserver: when the user clicks on
an element in the graphical view that contains source code location, a
message is sent from the webpage to the editor that contains a command
for letting the editor jump to the corresponding location in the
Dezyne model.  Note that large parts of the daemon and JavaScript as
part of a webpage rendered by the webbrowser is produced using Dezyne.

Dezyne currently has a state diagram view, a system view depicting the
composition of a system model out of its subcomponent, and as
mentioned before the sequence diagram for the interaction with the
simulator.  The user can either start the simulator directly, or
the simulator is started when the verification engine finds a problem
and generates a counter example. The next possible events are also
depicted in the sequence diagram and if the user selects a next event,
this request is send from the webpage to the daemon, and the daemon
reruns the simulator with the current trace of events extended with
the requested event.

\section{Application of Dezyne in Industry}

Dezyne and its predecessor ASD have been intensively used by several Verum
customers for more than a decade. Verum customers are mainly high tech
equipment manufacturers; their binding factor is that the
cost-of-non-quality is high in the market they operate.  Philips
Healthcare is one of the first customers and has the longest experience
in using Verum products.  For there initial projects the impact of
using this kind of low-entry Formal Methods based tooling has been
investigated \cite{eval0,eval1,eval2,eval3} and reported that the use
of the tool eliminated design errors earlier in the design process and
resulted in reduced development time and an ten fold reduction in
reported errors.

ASML is one of the leading manufacturers of semiconductor chip-making
equipment, and has been using our tooling in different projects for
different machines for some years now \cite{tabular, delete2017,
  mooreslaw}. They also reported a similar decrease in field defects
and decrease in development time.

The most striking difference between ASD and Dezyne is that ASD has
its own proprietary Microsoft Excel resembling editor and is as such
not language based. This has had all kind of implications for
incorporating ASD into existing software development toolchains. For
example, version management tooling is predominantly text based. This
alone was experienced as a substantial obstacle in the acceptance of
ASD. The feedback from users and the noted shortcomings have served as
primary input to design the Dezyne language and its tooling.

Dezyne is being used by other customers which names we cannot
disclose, operating in the field of semiconductor equipment industry,
electronic analytical instruments, and egg grading, packing and
processing machines.

\section{Conclusions and Future Directions}
Dezyne and its predecessor are being used at several high-tech
equipment manufacturing multinationals. So far several millions lines
of generated code are running in production around the world.

Software engineers without a formal methods background are productive
and comfortable due to the familiar concepts Dezyne is built on.
Furthermore the Dezyne language is designed to promote guiding
principles of software engineering. In order to avoid the state
explosion problem it is especially important to divide and conquer
complexity in manageable sub-components, by introducing internal
Dezyne interfaces that capture the responsibilities of these
sub-components. The Dezyne verification engine guides the user in
achieving an intuitive understanding of the complexity of the problem
at hand and provides the user with input as to which parts of the
system are candidates to be broken up.

The introduction of Dezyne and its predecessor has resulted in a
decrease in software development time, a vast reduction in integration
time, and a major decrease of reported field defects. This shows that
Formal Methods can bring a lot of added value to industrial software
engineers, provided that the method hides the
underlying complexities, can be used compositionally, generates
production quality and verified machine code,
is properly
packaged in a familiar IDEs and graphical interactive views, and
promotes the application of proper software engineering principles.

Currently, the Dezyne simulator is restricted to a single component
since this is used to debug the counter examples found by the
verification engine.  We are extending the simulator to bring multiple
components in scope and even complete systems. This will further
support the user in validating their systems, i.e. to check that the
intended use cases and user stories are present in the behaviour of
system.

We aim to extend the verification engine to prove properties on
several components combined. The challenge here is to find a proper
balance between the expressiveness of the language defining these
properties and the clarity of the expressed properties such that a
users can assess whether the formal properties capture the intended
informal requirements.

\nocite{*}
\bibliographystyle{eptcs}
\bibliography{generic}
\end{document}